\documentclass{PoS}
\usepackage{epsfig, rotating}

\PoS{PoS(LAT2005)193}

\dedicated{BI-TP-2005-34 and TKYNT-05-23 and BNL-NT-05-29}

\title{Heavy quark free energies and screening at finite temperature and density}

\ShortTitle{Heavy quark free energies and Debye screening masses at finite temperature and density}

\author{\speaker{Matthias D\"oring}\\
        Fakult\"at f\"ur Physik, Universit\"at Bielefeld, D-33615 Bielefeld, Germany\\
        E-mail: \email{doering@physik.uni-bielefeld.de}}

\author{S. Ejiri$^{\rm b}$, O. Kaczmarek$^{\rm a}$, F. Karsch$^{\rm a,c}$, E. Laermann$^{\rm a}$\\
$^{\rm a}$Fakult\"at f\"ur Physik, Universit\"at Bielefeld, D-33615 Bielefeld, Germany\\
$^{\rm b}$Department of Physics, The University of Tokyo, \small Tokyo 113-0033, Japan\\
$^{\rm c}$Physics Department, Brookhaven National Laboratory, Upton, NY 11973, USA}


\abstract{We study the free energies of heavy quarks calculated from Polyakov
  loop correlation functions in full 2-flavour QCD using the p4-improved
  staggered fermion action. A small but finite baryon number density is included
  via Taylor expansion of the fermion determinant in the baryo-chemical
  potential $\mu$. For temperatures above $T_c$ we extract Debye screening
  masses from the large distance behaviour of the free energies and compare their $\mu$-dependence to perturbative results.}

\FullConference{XXIIIrd International Symposium on Lattice Field Theory\\
                 25-30 July 2005\\
                 Trinity College, Dublin, Ireland}

\newcommand{\nc}[1]{\newcommand{#1}}
\nc{\bo}[1]{\mbox{\boldmath \( #1 \! \! \)  \unboldmath}}
\nc{\be}{\begin{eqnarray}}
\nc{\ee}{\end{eqnarray}}
\nc{\bew}{\begin{eqnarray*}}
\nc{\eew}{\end{eqnarray*}}
\nc{\f}[2]{\frac{#1}{#2}}
\nc{\suli}{\sum\limits}
\nc{\proli}{\prod\limits}
\nc{\nnn}{\nonumber \\}
\def\gsim{\raise0.3ex\hbox{$>$\kern-0.75em\raise-1.1ex\hbox{$\sim$}}}

\begin{document}

\section{Introduction}

Lattice QCD has widely been used to get information about the properties of
matter at high temperature and vanishing net baryon density $\mu_b$
\cite{review}. These studies have been extended to non-vanishing baryon
density. I.e., the equation of state has been discussed using Taylor expansion \cite{eos},
reweighting \cite{Fodor} or imaginary chemical potential techniques \cite{lombardo}.

Furthermore the free energy of static quark anti-quark sources at $\mu_b=0$
received much attention \cite{fthq}. We will now apply here the Taylor expansion
approach in order to extend the results on this free energy to non-zero quark
chemical potential. A first attempt to do this was discussed in
\cite{fodor-debye}. Furthermore we will analyze the large distance behaviour of
the free energy and determine the corresponding expansion coefficients of the screening mass, which can be compared to perturbative predictions.

We will restrict the discussion of the expansion up to the 2nd order in
$\mu_b$ here. Results up to the 6th order as well as furhter details on the
simlulation and the Taylor expansion technique can be found in \cite{fullpaper}.

\section{Taylor expansion method in QCD at finite density}

The staggered fermion partition function
\be
Z_\mu = \int DU \; \Delta(\mu) \;e^{-S}\;,
\ee 
where in our 2-flavour case $\Delta(\mu)$ is the square root of the fermion
determinant, can be Taylor expanded in powers of the quark chemical potential
$\mu = \mu_b/3$ by inserting the expansion of $\Delta(\mu)$ in powers of $\mu$.
\be
\Delta(\mu) = \Delta(0) \left( 1 + D_1 \mu + D_2 \mu^2 + \cdots \right)\;.
\ee
The odd orders in the expansion of $Z_\mu$ are vanishing because $D_n$ is imaginary for odd and real for
even $n$. The simulation can be done at $\mu = 0$, where the $D_n$ are
handled like observable quantities.

This procedure can be extended easily to the $\mu$-dependent
expectation value of an observable ${\cal A}$ that does not directly depend on
$\mu$,
\be
\left< {\cal A} \right>_\mu &=& \f{1}{Z_\mu} \int DU \;{\cal A}
\Delta(\mu)\;e^{-S} = \f{\left< {\cal A} \cdot 1\right>_0 + \left<{\cal A} D_1
  \right>_0 \mu + \cdots + \left< {\cal A}
    D_6 \right>_0 \mu^6}{1 + \left< D_2 \right>_0 \mu^2
      +\left< D_4 \right>_0 \mu^4 + \left< D_6 \right>_0 \mu^6} + {\cal O}(\mu^
7)\;,\nnn
&=& \left< {\cal A} \right>_0 \left( 1 + a_1 \mu + a_2 \mu^2 + \cdots +  a_6
  \mu^6 \right) + {\cal O}(\mu^7)\;.
\ee
We apply this scheme to the Polyakov loop correlation functions discussed
in the next section.

\section{Heavy quark free energies}

\begin{figure}[t]
\bew
\begin{array}{cc}
\begin{turn}{270} \epsfig{file=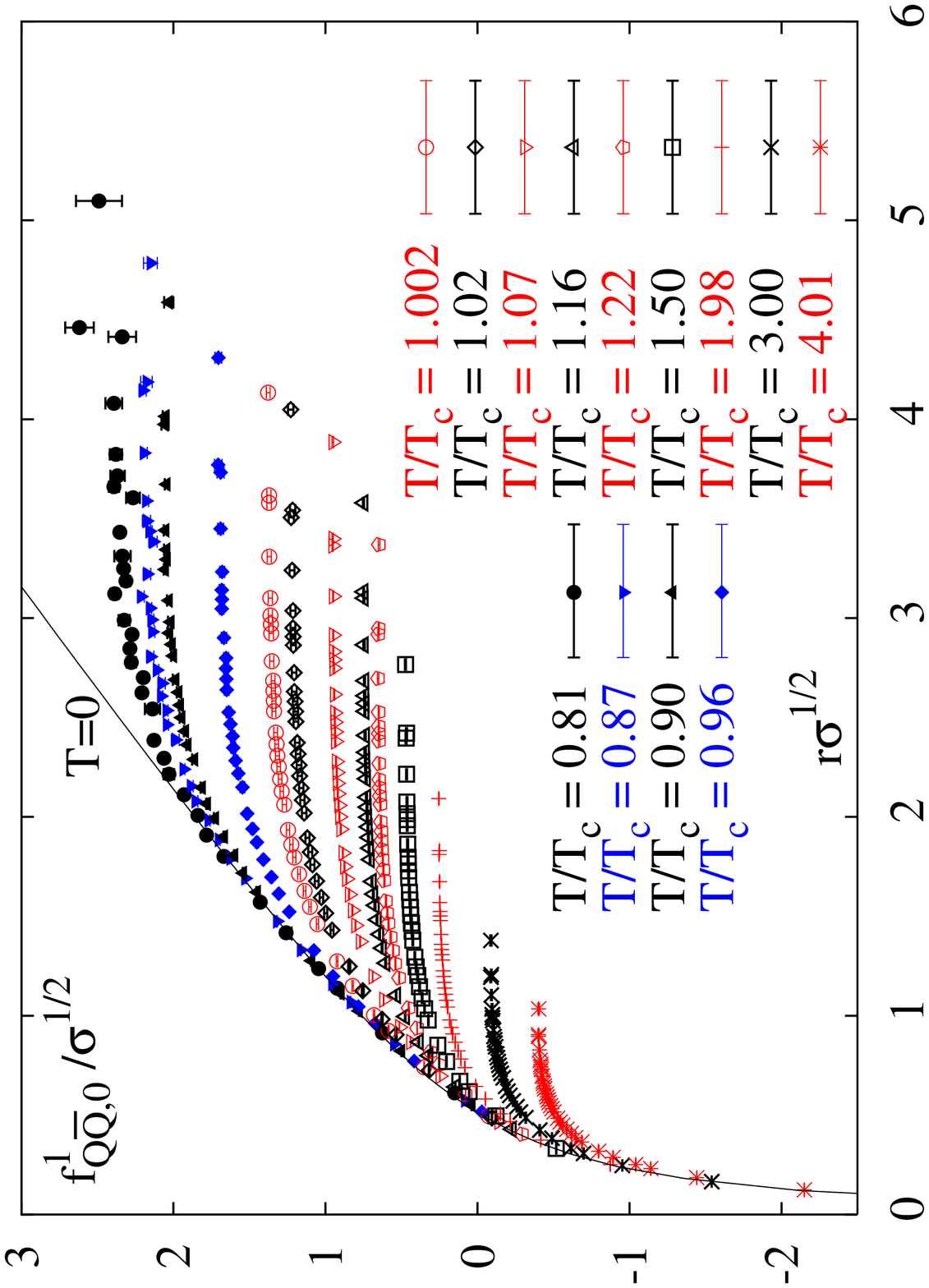, width=0.35\textwidth} \end{turn} &
\begin{turn}{270} \epsfig{file=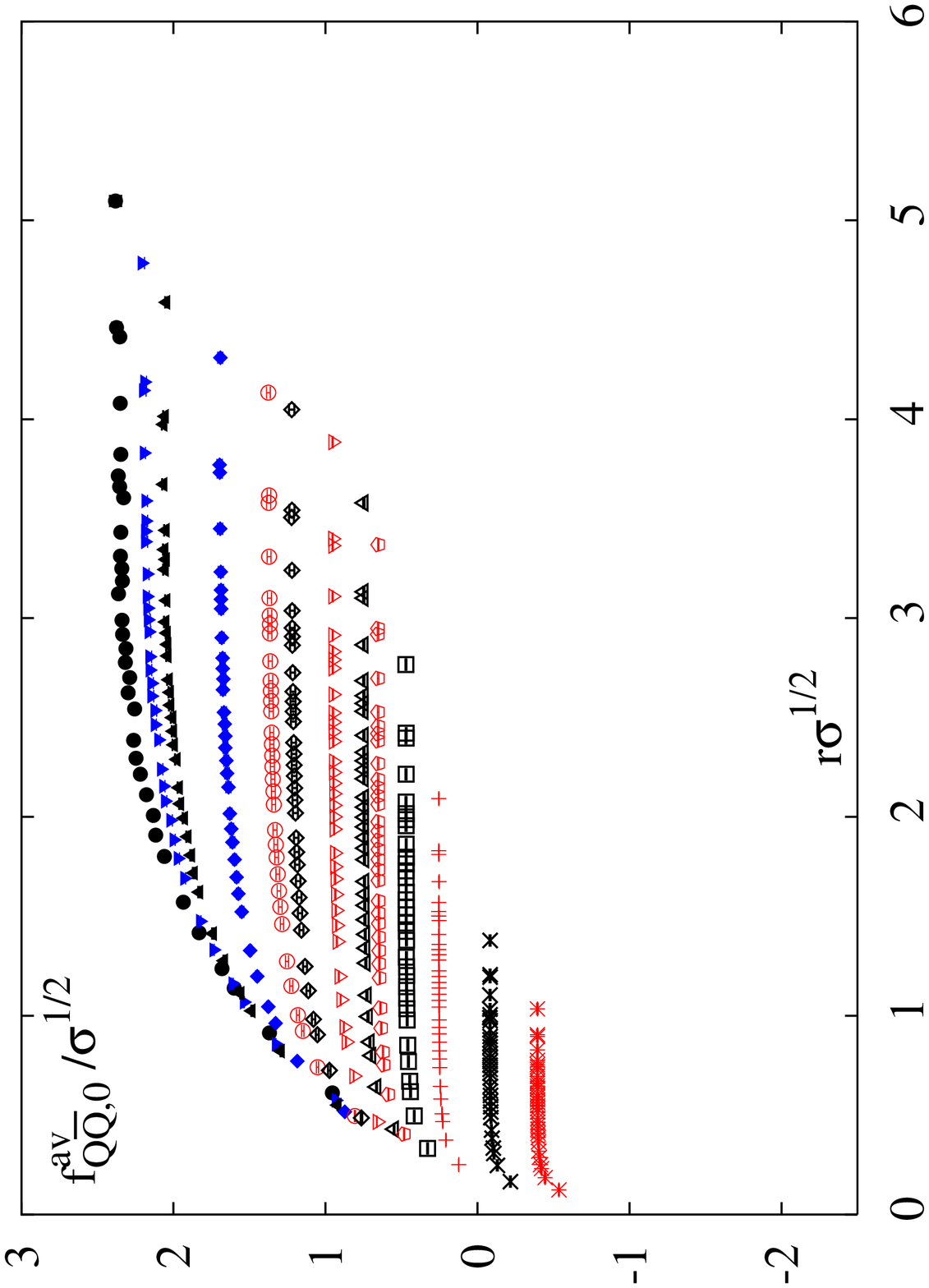, width=0.35\textwidth} \end{turn} \\
\mbox{(a)} & \mbox{(b)}\\
\end{array}\\
\eew
\caption{The $0^{th}$ order coefficients $f^1_{Q\bar{Q},0}$ and 
$f^{\rm av}_{Q\bar{Q},0}$ for the singlet (a) and colour averaged (b) free energies. $f^1_{Q\bar{Q},0}$ is matched to the $T=0$ heavy quark potential at small distances (a). \label{order0}}
\end{figure}

\begin{figure}[t]
\bew
\begin{array}{cc}
\begin{turn}{270} \epsfig{file=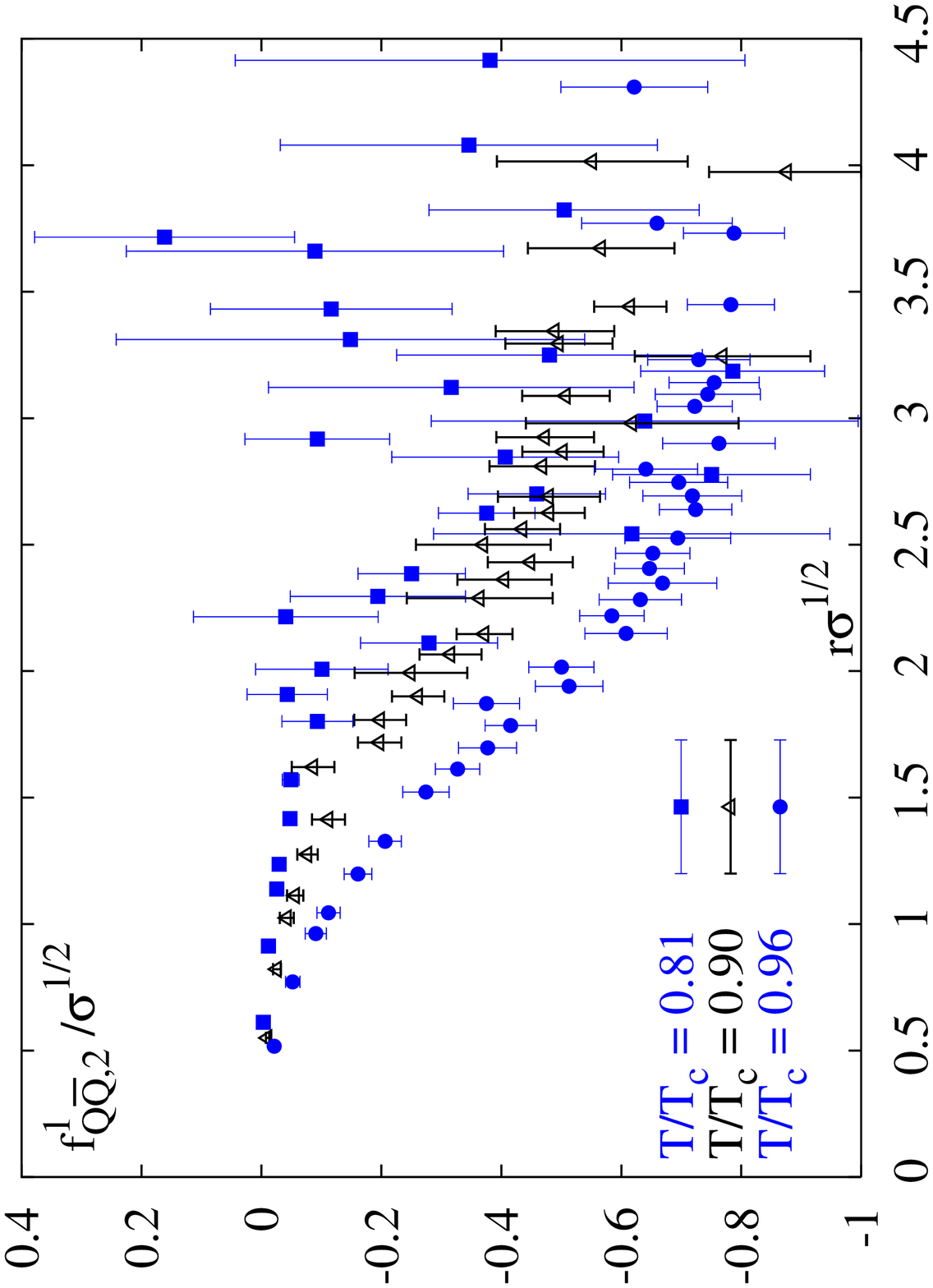, width=0.35\textwidth} \end{turn} &
\begin{turn}{270} \epsfig{file=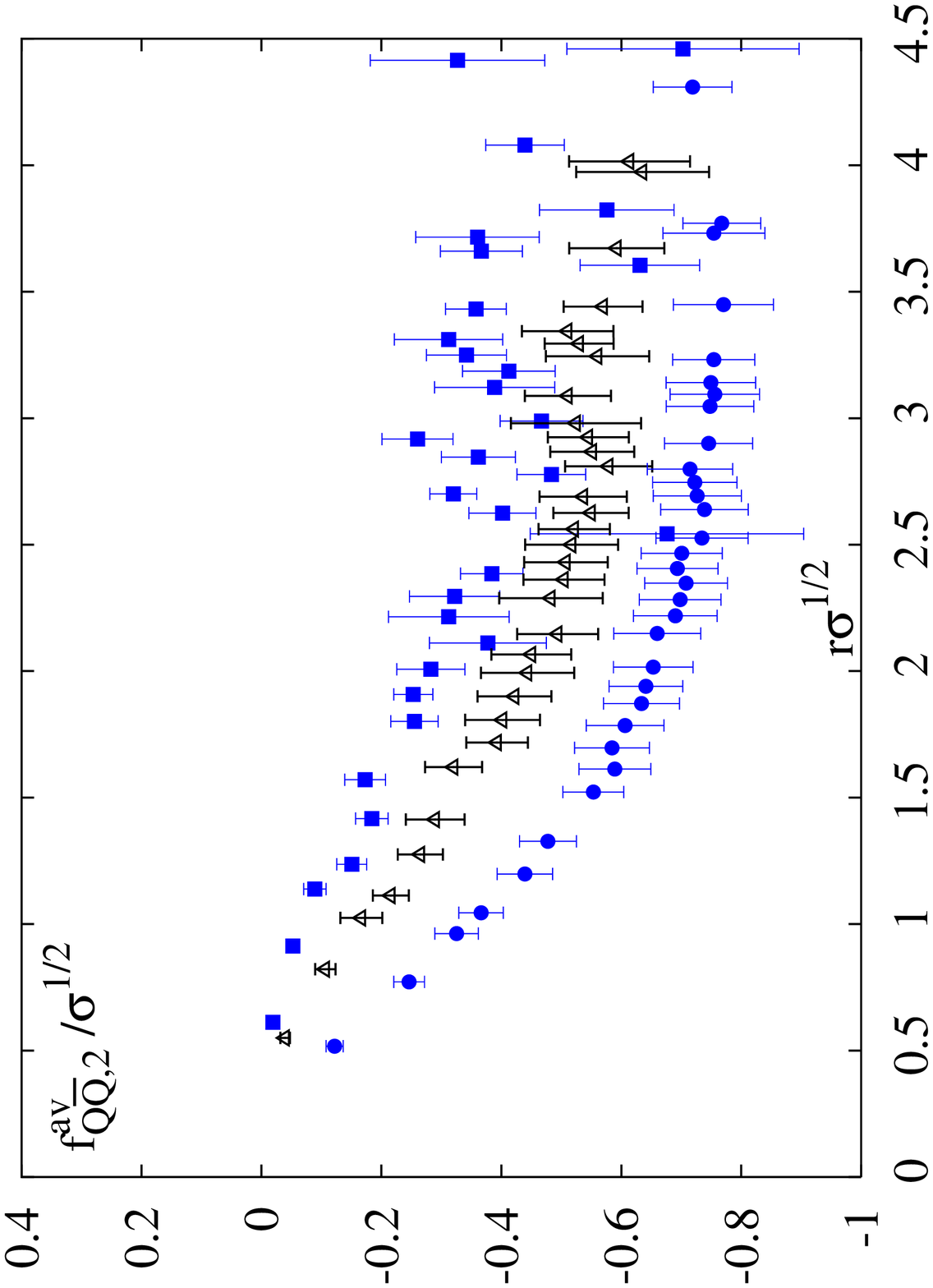, width=0.35\textwidth} \end{turn} \\
\mbox{(a)} & \mbox{(b)}\\
\begin{turn}{270} \epsfig{file=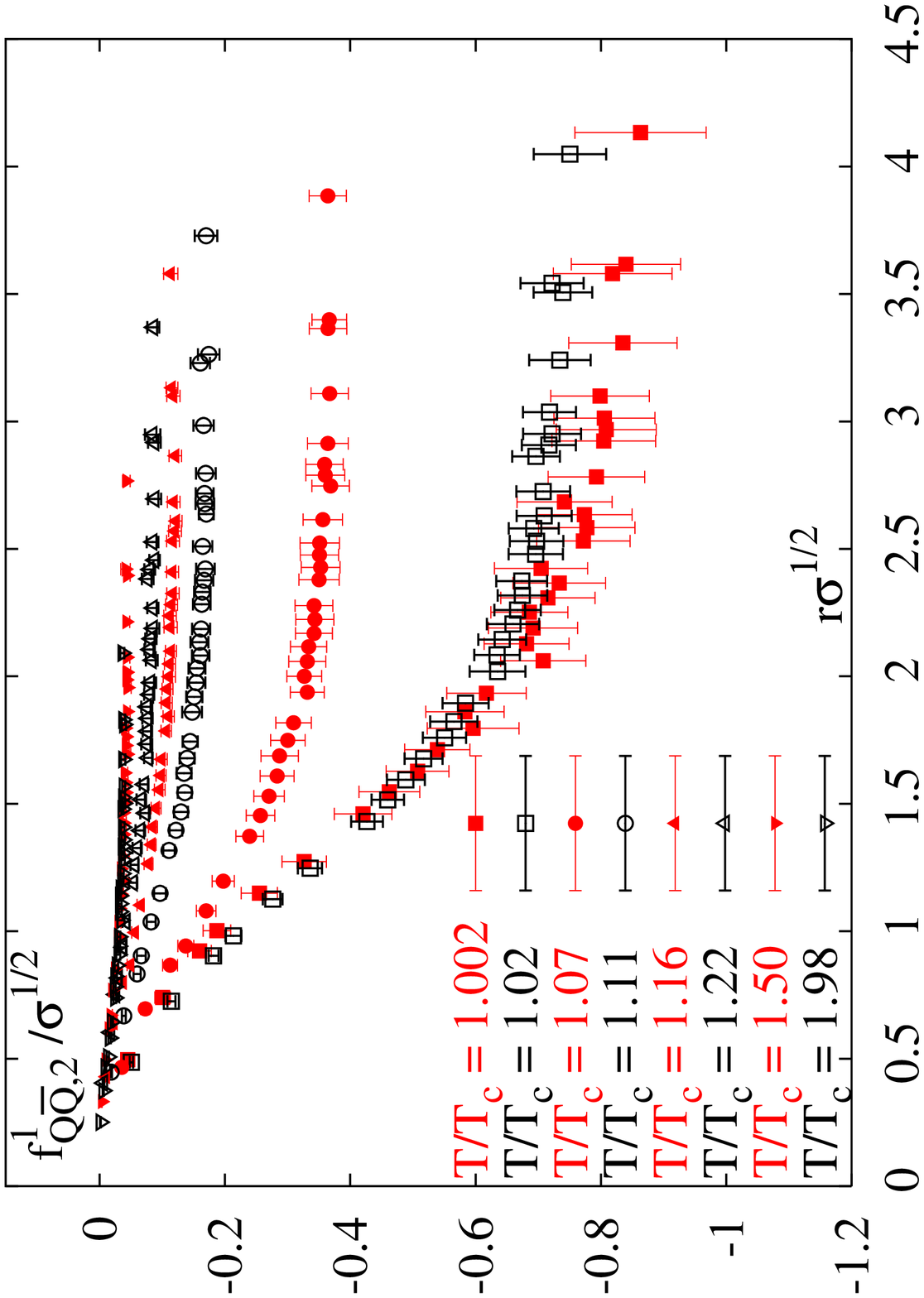, width=0.35\textwidth} \end{turn} &
\begin{turn}{270} \epsfig{file=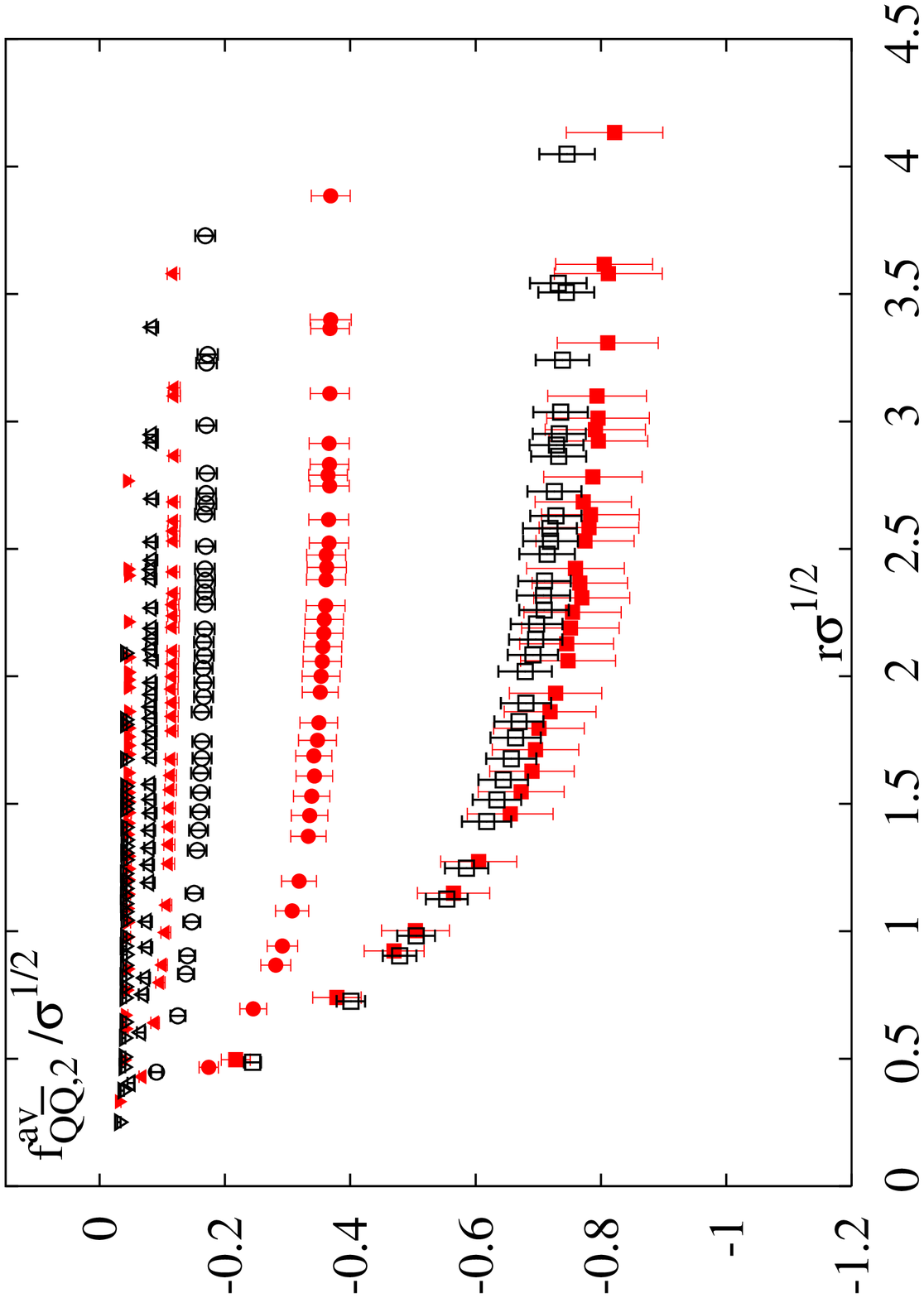, width=0.35\textwidth} \end{turn} \\
\mbox{(c)} & \mbox{(d)}
\end{array}\\
\eew
\caption{The $2^{nd}$ order coefficients of the singlet (a,c) and colour 
averaged (b,d) free energies below (a,b) and above (c,d) $T_c$. \label{order2}}
\end{figure}

A heavy (static) quark $Q$ at site $\bo{x}$ is represented by the Polyakov 
loop, 
\be
L(\bo{x}) = \proli_{x_4 = 1}^{N_\tau} U_4(\bo{x}, x_4)\;, 
\ee
which is an SU(3) matrix. A heavy anti-quark $\bar{Q}$ is described by the
corresponding hermitian conjugate matrix. The colour averaged and singlet
$Q\bar{Q}$-free energies $F^{\rm av}_{Q\bar{Q}}(r, T, \mu)$ and $F^{\rm
  1}_{Q\bar{Q}}(r, T, \mu)$ can be calculated from the corresponding correlation
functions ${\cal O}^{\rm av,1}(r)$,
\be
F^{\rm av,1}_{Q\bar{Q}}(r, T, \mu) = -T \; \ln{\left< {\cal O}^{\rm
      av,1}(r)\right>} = f^{av,1}_{Q\bar{Q},0}(r, T) + f^{av,1}_{Q\bar{Q},2}(r,
T) \left(\f{\mu}{T}\right)^2 + {\cal O}\left(\left(\f{\mu}{T}\right)^4\right)\;,
\ee
where
\be
{\cal O}^{\rm av}(r) &=& \f{1}{{\cal N}} \, \f{1}{N_c^2} \, \suli_{\bo{x},
  \bo{y}} \mbox{Tr} L(\bo{x}) \, \mbox{Tr} L^\dagger(\bo{y})\;, \nonumber \\
{\cal O}^{\rm 1}(r) &=& \f{1}{{\cal N}} \, \f{1}{N_c} \, \suli_{\bo{x}, \bo{y}}
\mbox{Tr} L(\bo{x})  L^\dagger(\bo{y})\;.
\ee
In Fig.~\ref{order0} and ~\ref{order2} we show the results for the 0th and 2nd order
expansion coefficients in $\mu/T$ for temperatures above and below $T_c$. Because $f^{\rm av, 1}_{Q\bar{Q}, 2}(r, T)$ is always negative,
we find that for small $\mu$ the free energy of a static quark anti-quark pair
decreases in a medium with a net excess of quarks or anti-quarks.

\begin{figure}[t]
\bew
\begin{array}{cc}
\begin{turn}{270} \epsfig{file=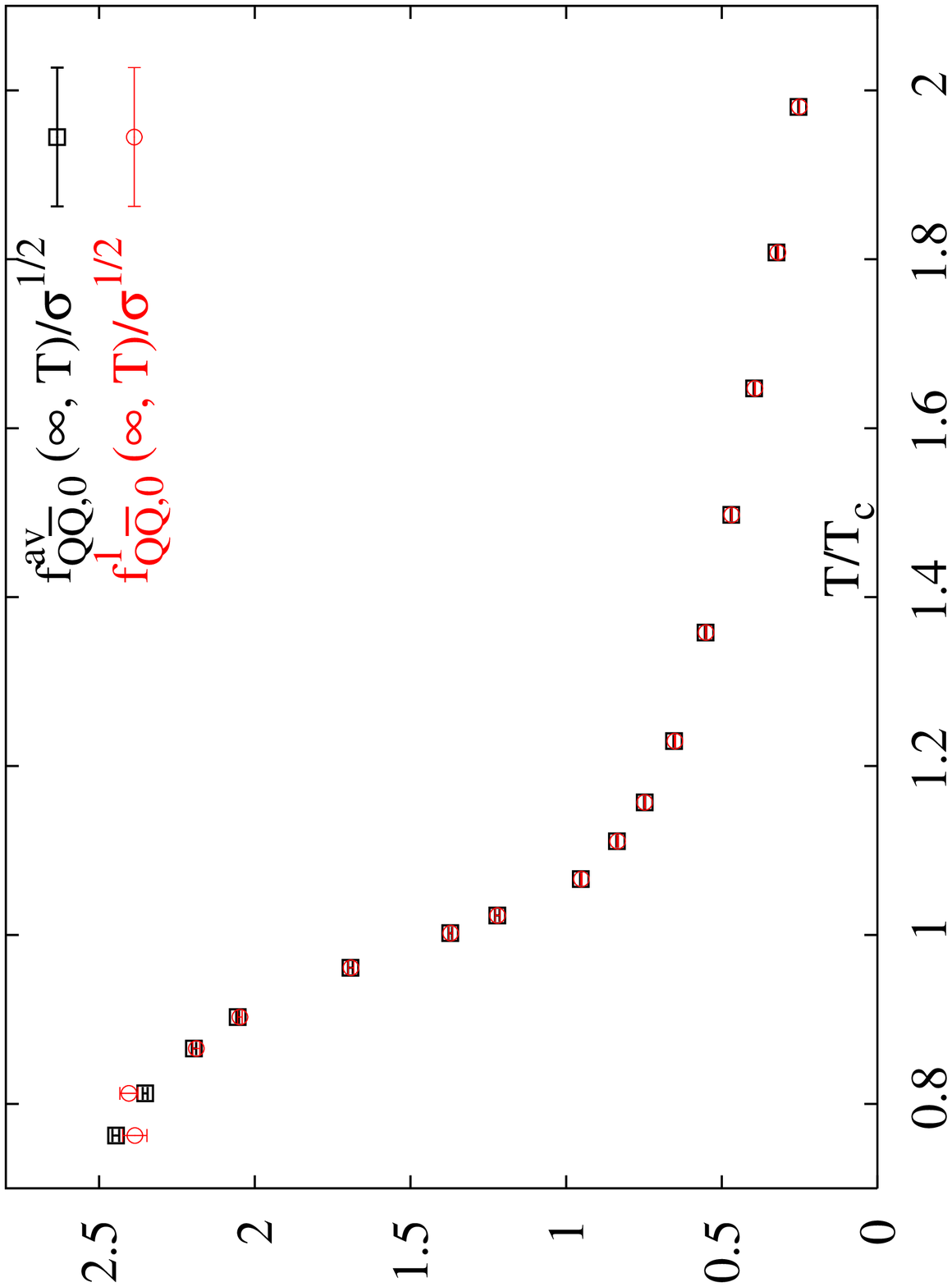, width=0.35\textwidth} \end{turn} &
\begin{turn}{270} \epsfig{file=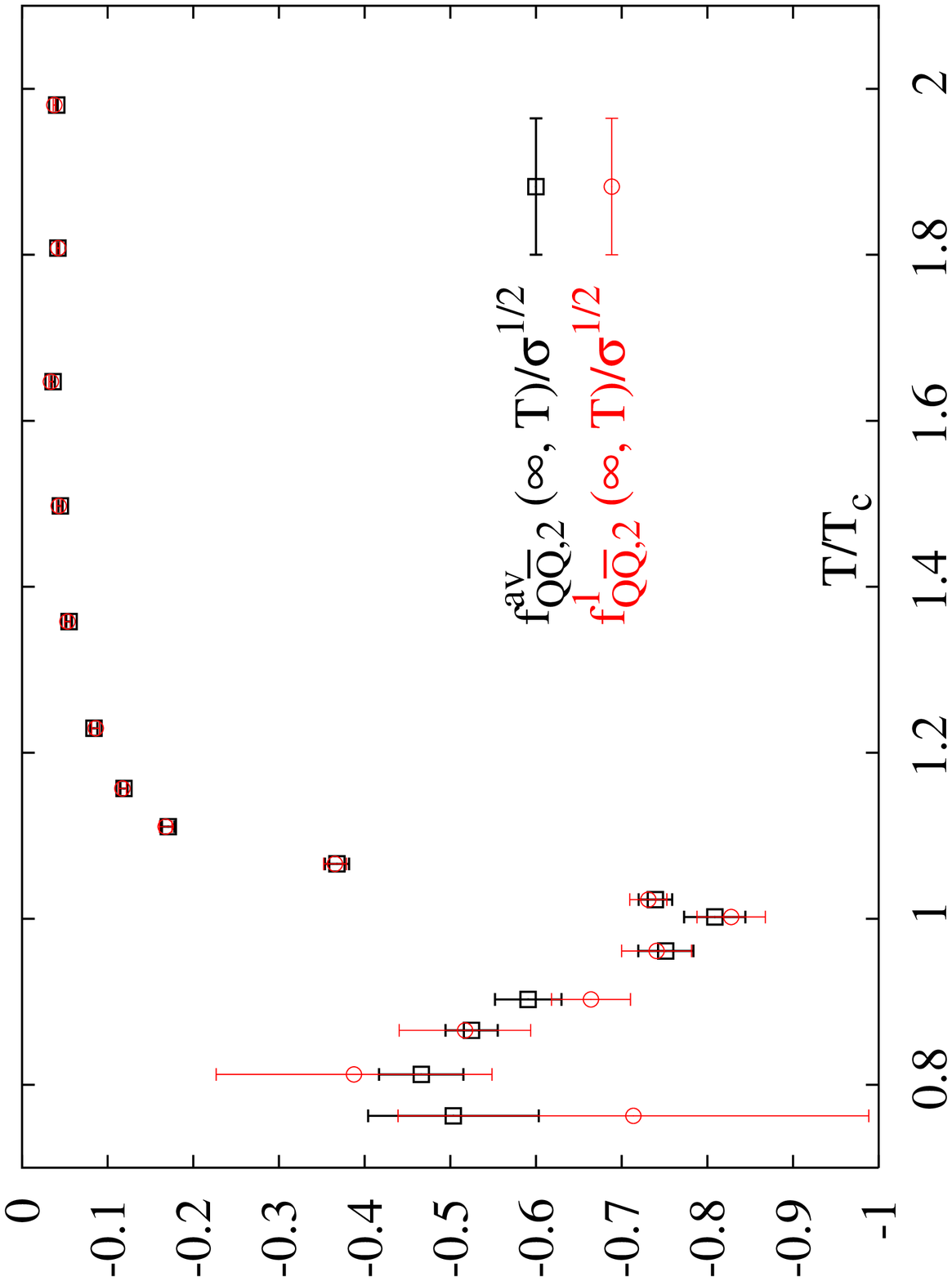, width=0.35\textwidth} \end{turn} \\
\mbox{(a)} & \mbox{(b)}
\end{array}
\eew
\caption{The coefficients for the singlet and colour averaged free energies 
at infinite distance $rT$ versus temperature.\label{coeffrt}}
\end{figure}

We determine the large distance value of these coefficients, shown in
Fig.~\ref{coeffrt}, by taking the weighted average of the values at the five
largest distances. $f^{\rm av,1}_{Q\bar{Q}, 2}(\infty, T)$ signs the transition
temperature by showing a pronounced peak at $T_c$. The coefficients of the
colour averaged and singlet free energy are identical at infinite distance.

\section{Screening masses}

\begin{figure}[t]
\bew
\begin{array}{cc}
\begin{turn}{270} \epsfig{file=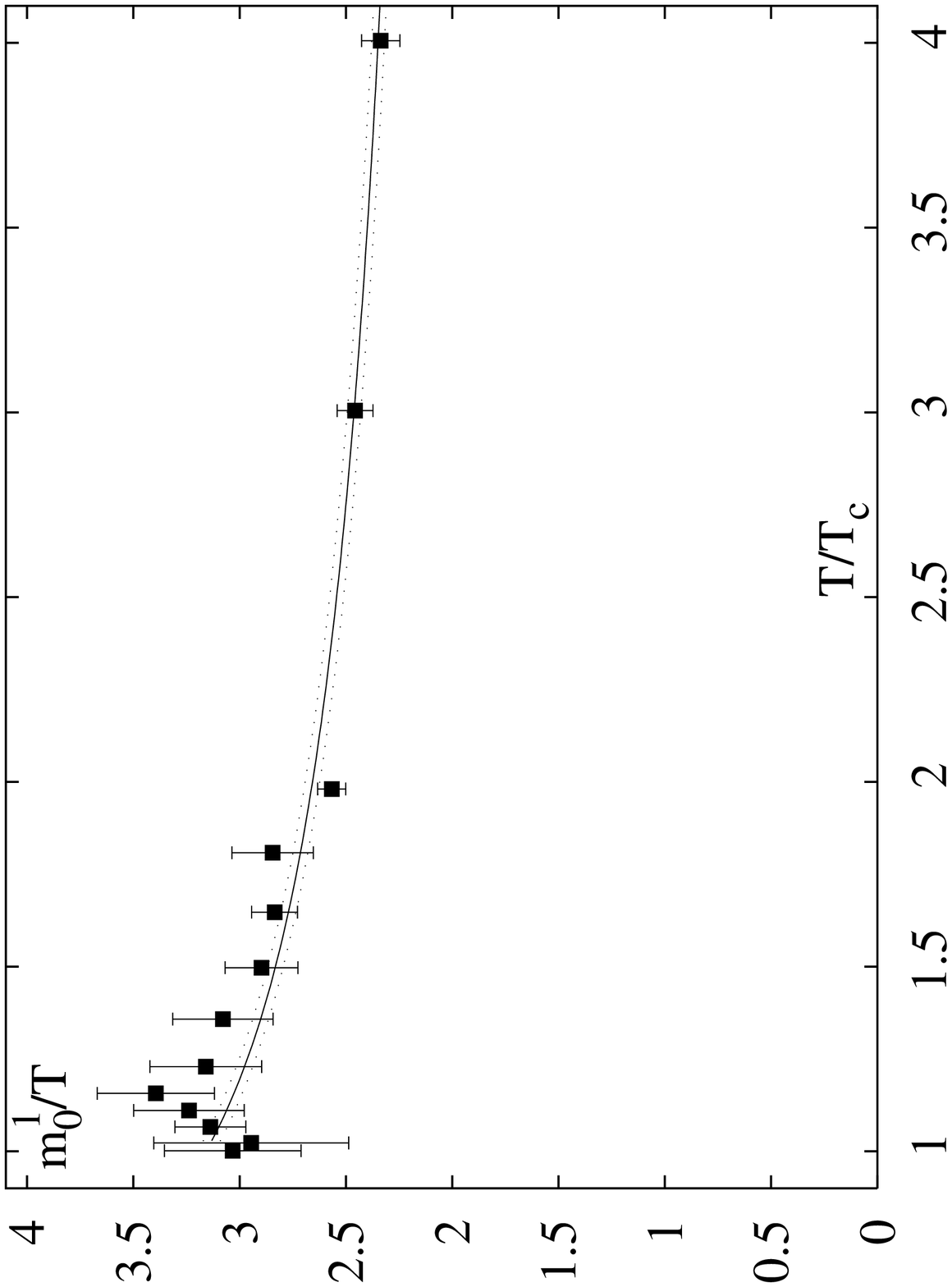, width=0.35\textwidth}   \end{turn} &
\begin{turn}{270} \epsfig{file=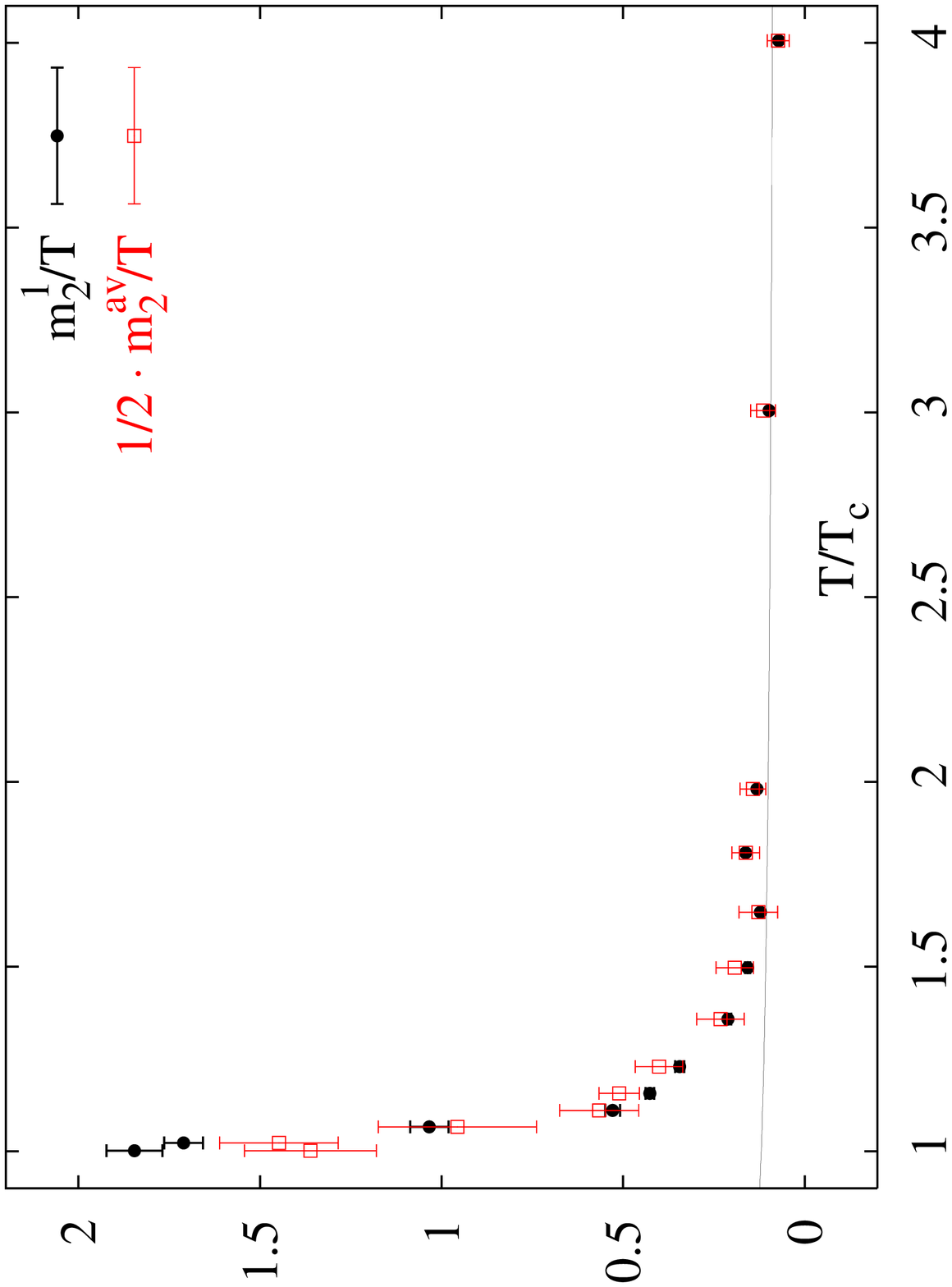, width=0.35\textwidth} \end{turn}\\
\mbox{(a)} & \mbox{(b)}\\
\end{array}
\eew
\caption{0th and 2nd order expansion coefficients of the screening
  mass. The solid lines are connected to the LO perturbation theory.\label{mass2}}
\end{figure}

For temperatures above $T_c$ and large distances $r$ the heavy quark
free energies are expected to be screened,
\begin{eqnarray}
\Delta F^{\rm av,1}_{Q\bar{Q}}(r, T, \mu) &=&
F^{\rm av,1}_{Q\bar{Q}}(\infty, T, \mu) - F^{\rm av,1}_{Q\bar{Q}}(r, T,\mu) \sim
\f{1}{r^n}e^{-m^{\rm av,1}(T,\mu) r} \label{delta}
\end{eqnarray}
with $n = 1, 2$ for the singlet and colour averaged free energies respectively.
In the infinite distance limit we thus get,
\be
m^{\rm av,1}(T,\mu) = -\lim_{r\rightarrow \infty} \frac{1}{r} \ln \left( \Delta F^{\rm av,1}_{Q\bar{Q}}(r, T,\mu) \right)\; . \label{debye-ansatz}
\ee
Expanding the logarithm in powers of $\mu$ we see that only the even orders of
the expansion coefficients for the screening mass are non-zero, i.e. the second order coefficient can be
written as
\be
m_2^{av,1}(T) = -\lim_{r\rightarrow \infty} \f{1}{r} \frac{\Delta
  f^{av,1}_{Q\bar{Q},2}(r, T)}{\Delta f^{av,1}_{Q\bar{Q},0}(r, T)} \;, \label{mass22}
\ee
where we write $\Delta f^{{\rm av},1}_{Q\bar{Q},n}$ for the expansion
coefficients of $\Delta F^{\rm av,1}_{Q\bar{Q}}$. We use this expression to
determine the expansion coefficient of order 2 from the infinite distance
limit. For the 0th order coefficient we refer to fits of the form of
eq.~\ref{delta}. In fact the rational expression under the limit in
eq.~\ref{mass22} shows only little $r$-dependence and we see a wide plateau
starting close to $rT = 0$. We therefore determine $m^{1,av}_2(T)$ by fitting
the ratio in eq.~\ref{mass22} to a constant in the plateau range. The
resulting coefficient, shown in Fig.~\ref{mass2} (b) is always positive and
leads to an enhancment of the screening mass for small $\mu_b \ne 0$. At high
temperatures $m^{1,av}_2(T)$ is in remarkable agreement with the LO high
temperature perturbative prediction resulting from the $\mu$-expansion of the Debye mass
\be
m^D(T, \mu) = g(T) T \sqrt{\f{N_c}{3} + \f{N_f}{6} + \f{N_f}{2 \pi^2} \left(
    \f{\mu}{T} \right)^2}\;,
\ee
where we introduce an additional scale factor $A$ via $m^1(T) = A \cdot m^D(T)$,
which is determined from fitting $m^1_0(T)$. We find $A=1.397(18)$. Finally we see, that $m^1_2(T)$ and
$m^{av}_2(T)$ differ by a factor $1/2$, which is expected at high temperature
for all the coefficients but has not been observed for the 0th order so far.

\section{Conclusions}

We analyzed the dependence of heavy quark free energies and screening on the
baryon-chemical potential. The free energies are decreasing to first order in
$\mu^2$, while the screening mass is increasing. This suggests that the
screening length in a baryon or anti-baryon rich quark gluon plasma decreases
with increasing value of the chemical potential, which is consistent with the
expectation that a non-zero $\mu$ shifts the transition to lower
temperatures. The screening behaviour is in good agreement with perturbation
theory for $T/T_c \gsim 2$. We observed that the $\mu$-dependent corrections of
the colour averaged screening mass are twice as large as those of the colour
singlet. This suggests that the contribution to the $\mu$-dependent corrections
of the colour averaged screening mass is due to two-gluon exchange.

\section{Acknowledgments}
\label{ackn}
Our work has been supported partly through the DFG under grant KA 1198/6-4, the
GSI collaboration grant BI-KAR and a grant of the BMBF under contract no. 06BI106. MD is supported through a fellowship of the DFG funded graduate school GRK 881. 
The work of FK has been partly supported by a contract DE-AC02-98CH1-886 with
the U.S. Department of Energy.


\begin{thebibliography}{99}

\bibitem{review} F.~Karsch and E.~Laermann, hep-lat/0305025.

\bibitem{eos} C.~R.~Allton, S.~Ejiri, S.~J.~Hands, O.~Kaczmarek, F.~Karsch,
E.~Laermann and C.~Schmidt, Phys.\ Rev.\ D {\bf 68} (2003) 014507;\\
C.~R.~Allton, M.~Doring, S.~Ejiri, S.J.~Hands, O.~Kaczmarek,
F.~Karsch, E.~Laermann, K.~Redlich, Phys. Rev D 71 (2005) 054508.
\bibitem{Fodor} Z.~Fodor, S.~D.~Katz and K.~K.~Szabo, Phys.\ Lett.\ B {\bf 568} (2003) 73;\\
F.~Csikor, G.~I.~Egri, Z.~Fodor, S.~D.~Katz, K.~K.~Szabo and A.~I.~T\'oth, JHEP {\bf 0405} (2004) 046.
\bibitem{lombardo} M.~D'Elia and M.~P.~Lombardo, Phys.\ Rev.\ D {\bf 70} (2004) 074509.
\bibitem{fthq} O.~Kaczmarek, F.~Karsch, F.~Zantow, P.~Petrecky, Phys.Rev. {\bf D 70} (2004) 074505.
\bibitem{fodor-debye}
Z.~Fodor, S.~D.~Katz, K.~K.~Szabo and A.~I.~T\'oth, Nucl.\ Phys.\ B (Proc.\ Suppl.)\ {\bf 140} (2005) 508.
\bibitem{fullpaper} M.~D\"oring, S.~Ejiri, O.~Kaczmarek, F.~Karsch, E. Laermann,
  hep-lat/0509001.


\end{thebibliography}
\end{document}